\documentclass[conference]{IEEEtran}
\IEEEoverridecommandlockouts
\usepackage{cite}
\usepackage{graphicx}
\usepackage{epstopdf}
\usepackage{amsmath,amssymb,amsfonts}
\usepackage{algorithmic}
\usepackage{textcomp}
\usepackage{xcolor}
\usepackage[ruled,lined,linesnumbered]{algorithm2e}
\usepackage{makecell}
\usepackage{extarrows}
\usepackage{multicol}
\usepackage{subcaption}
\usepackage{multirow}
\usepackage{bm}
\usepackage{colortbl}
\usepackage{amsthm}
\usepackage{mathrsfs}
\usepackage{booktabs}
\usepackage{siunitx}
\usepackage{tabularx,booktabs}
\usepackage[flushleft]{threeparttable}
\usepackage[short]{optidef}

\usepackage{xcolor}
\usepackage{xparse}
\usepackage[normalem]{ulem}
\usepackage[most]{tcolorbox}
\definecolor{promptframe}{HTML}{2C3E50}
\definecolor{promptback}{HTML}{F4F6F8}
\definecolor{promptslot}{HTML}{B22222}
\newtcolorbox{promptbox}[1][]{enhanced, breakable, sharp corners,
  colback=promptback, colframe=promptframe, boxrule=0.4pt, left=6pt, right=6pt,
  top=4pt, bottom=4pt, fonttitle=\bfseries\sffamily\small,
  coltitle=white, colbacktitle=promptframe,
  attach boxed title to top left={xshift=6pt, yshift=-2pt},
  boxed title style={sharp corners, boxrule=0pt, top=1pt, bottom=1pt,
    left=4pt, right=4pt}, #1}

\newif\ifshownotes
\shownotesfalse     

\newcommand{\defineauthor}[3]{%
  \expandafter\gdef\csname author@#1@name\endcsname{#2}%
  \expandafter\gdef\csname author@#1@color\endcsname{#3}%
}

\NewDocumentCommand{\note}{m m}{%
  \ifshownotes
    {\textbf{[\csname author@#1@name\endcsname: #2]}}
  \fi
}

\defineauthor{GLM5}{GLM-5}{purple}
\defineauthor{SK}{Codex}{blue}
\defineauthor{SP}{Supervisor}{red}

\definecolor{proposedblue}{RGB}{242,242,255}     
\definecolor{proposedblueB}{RGB}{255, 220, 225}   

\def\BibTeX{{\rm B\kern-.05em{\sc i\kern-.025em b}\kern-.08em
T\kern-.1667em\lower.7ex\hbox{E}\kern-.125emX}}

\begin{document}

\title{When Eavesdroppers Reason: Agentic Eavesdropping Attacks on Semantic Communication}
\author{
\IEEEauthorblockN{
Shunpu Tang\IEEEauthorrefmark{2}, Qianqian Yang\IEEEauthorrefmark{2}, Zhiguo Shi\IEEEauthorrefmark{2}, 
Jiming Chen\IEEEauthorrefmark{3}\IEEEauthorrefmark{4}, and Xuemin (Sherman) Shen\IEEEauthorrefmark{5}
    }
\IEEEauthorblockA{ 
  \IEEEauthorrefmark{2} College of Information Science and Electronic Engineering, Zhejiang University, Hangzhou, China \\
  \IEEEauthorrefmark{3} State Key Laboratory of Industrial Control Technology, Zhejiang University, Hangzhou, China \\
  \IEEEauthorrefmark{4} School of Automation, Hangzhou Dianzi University, Hangzhou, China \\
  \IEEEauthorrefmark{5}  Department of Electrical and Computer Engineering, University of Waterloo, Waterloo,  Canada  \\
Email: \{tangshunpu,  qianqianyang20, shizg, cjm\}@zju.edu.cn, sshen@uwaterloo.ca
}
\thanks{This work is partly supported by the NSFC under grant No. 62293481 and No. 62571487, by the National Key R\&D Program of China under Grant 2024YFE0200802, and by the Zhejiang Provincial Natural Science Foundation of China under Grant No. LZ25F010001.  (Corresponding author: Qianqian Yang.)}
}

\maketitle

\thispagestyle{empty}
\pagestyle{empty}
\begin{abstract}
  Semantic communication (SemCom) has emerged as a promising paradigm for next-generation networks. However, its typical end-to-end joint source--channel coding (JSCC) architecture also raises serious privacy concerns. To guide future secure SemCom design, it is important to understand how serious such leakage can be. Nevertheless, existing eavesdropping attacks mainly rely on fixed-configuration solvers and often require instantaneous wiretap channel state information (CSI) to achieve effective privacy inference. This may lead future secure SemCom designs to overlook potentially severe risks. To address this, we propose a large language model (LLM)-orchestrated agentic eavesdropper. Specifically, the proposed eavesdropper forms a closed-loop workflow with three functional agents. The optimization agent adaptively performs joint semantic-and-channel inversion to recover private information from the intercepted signal without requiring wiretap CSI. The perception agent evaluates the effectiveness of the optimization agent and assesses whether the recovered private semantics are reasonable, providing feedback to the optimization agent. The refinement agent further analyzes the recovered content and uses a generative prior to refine promising candidates into more realistic and complete private reconstructions while preserving consistency with the intercepted signal. Simulation results over a MIMO Rayleigh fading channel show that the proposed eavesdropper achieves more than $75\%$ eavesdropping success rate at $\mathrm{SNR}\geq 5$~dB even without wiretap CSI,     highlighting a severe privacy threat that future secure SemCom systems must address.
\end{abstract}


\section{Introduction}
Semantic communication (SemCom) has emerged as a promising paradigm for upcoming sixth-generation (6G) networks, in which the transmitter delivers only task-relevant semantic information rather than raw bits to the receiver~\cite{Semantic1}. Compared with traditional digital communication systems, SemCom can significantly reduce communication overhead and improve robustness under low signal-to-noise ratio (SNR) conditions. However, this advantage may also benefit eavesdroppers, leading to serious privacy leakage. This is because the typical joint source--channel coding (JSCC) architecture commonly used in SemCom learns source compression and channel coding in an end-to-end manner, often without quantization, making bit-level encryption difficult to apply directly. Moreover, conventional physical-layer security (PLS) methods, which typically rely on channel advantages between legitimate and eavesdropping links, may be insufficient, since an eavesdropper can still infer sensitive information even under a much worse channel condition~\cite{Semantic_security_zhaohui}.

To verify this security vulnerability, prior works have studied detailed eavesdropping attacks against SemCom systems. Specifically, the authors in~\cite{Semantic_security_maojun} showed that an eavesdropper equipped with a simple decoder can recover recognizable source information even at low SNR, such as $0$~dB. The authors in~\cite{Semantic_security_yuhao} proposed a model inversion attack (MIA), where the eavesdropper only knows the semantic encoder and reconstructs the source data by optimizing an input whose encoded representation matches the intercepted signal. More recently, the work in~\cite{tang2025towards} extended this threat model to an intelligent eavesdropper that exploits generative priors to produce more realistic reconstructions, further increasing the risk of privacy leakage. However, most existing attacks rely on handcrafted solver configurations for specific channel conditions and assume that the eavesdropper has access to instantaneous wiretap CSI to achieve effective leakage inference. Therefore, it remains unclear whether a CSI-agnostic eavesdropper can still cause severe privacy leakage in practical scenarios where Eve only knows the statistical distribution of the wiretap channel  ~\cite{Full-Duplex_Radio_for_Securing_Wireless_Network,Secure_MISO_Wiretap_Channels}. As a result, secure SemCom designs developed against existing attack strategies may overlook this practical threat, leaving both the corresponding security analysis and defense design insufficiently explored.

\begin{figure*}[t!]
  \centering
  \includegraphics[width=0.8\textwidth]{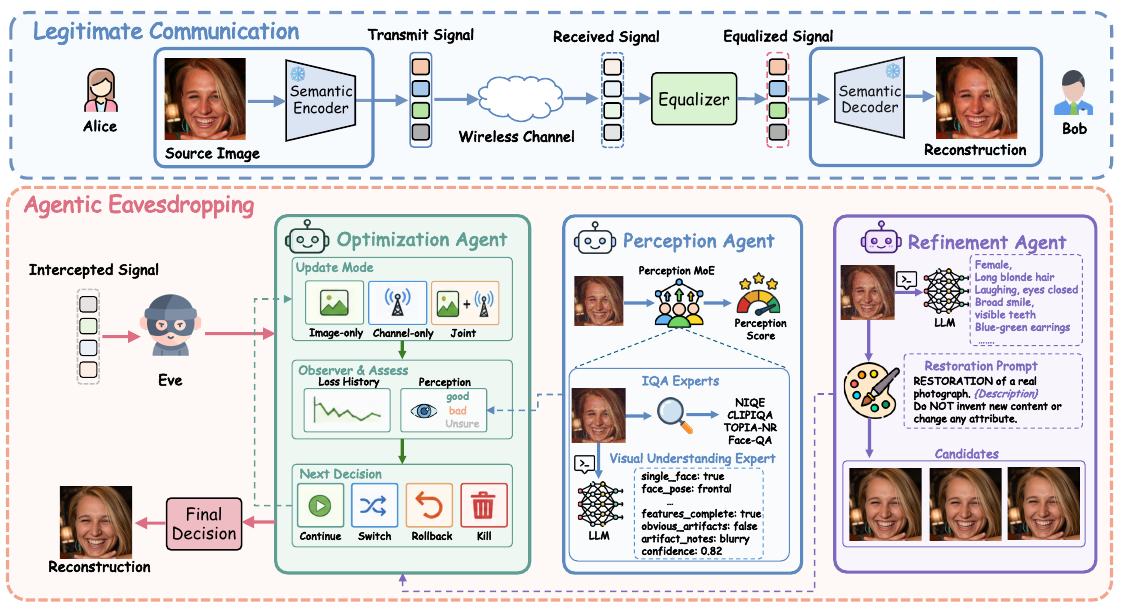}
  \caption{Illustration of the proposed agentic eavesdropper framework, where there are three agents: the optimization agent, the perception agent, and the refinement agent, working together in a sequential decision-making framework to reconstruct the source image from the intercepted signal.}
  \label{fig:agentic-eavesdropping}
\end{figure*}

In parallel, recent advances in large language models (LLMs) have shown that LLMs can act as autonomous agents capable of perception, reasoning, planning, and tool use~\cite{yao2023react, yao2023tree,llm-as-a-judge}. The wireless communication community has also begun to explore agentic AI for network optimization, resource management, and protocol design~\cite{zhang2025toward}. These developments motivate us to revisit SemCom eavesdropping from an agentic perspective and ask: \emph{Can an LLM-empowered eavesdropper further increase privacy leakage by analyzing intercepted signals, planning attacks, and using external tools to improve recovery?}

To answer this question and provide insights for future secure SemCom design, we propose an agentic eavesdropper that reformulates the eavesdropping attack as an LLM-driven sequential decision process. Specifically, the proposed framework forms a closed loop of three specialized agents. The optimization agent performs short-step joint semantic-and-channel inversion within resumable sessions. The perception agent evaluates inversion quality and semantic plausibility with no-reference quality assessment tools and a multimodal LLM, and provides feedback for subsequent optimization. The refinement agent analyzes promising candidates and uses an external generative prior to improve their realism and semantic completeness while maintaining consistency with the intercepted signal. Simulations on the FFHQ dataset over a MIMO Rayleigh fading channel show that the proposed agentic eavesdropper achieves more than $75\%$ eavesdropping success rate at $\mathrm{SNR}\geq 5$~dB without wiretap CSI, outperforming conventional glass-box MIA baselines.

\section{System Model}
\subsection{Transmission Model}
We consider a SemCom system for wireless image transmission over a multiple-input multiple-output (MIMO) fading channel, where the transmitter is equipped with $N_{\text{t}}$ antennas and the receiver with $N_{\text{r}}$ antennas. Let the source image be $\bm{x} \in \mathbb{R}^{3 \times N_{\text{H}} \times N_{\text{W}}}$, where the first dimension corresponds to the RGB color channels, and $N_{\text{H}}$ and $N_{\text{W}}$ denote the image height and width, respectively. Let $N = 3 \times N_{\text{H}} \times N_{\text{W}}$ denote the source dimension and $T$ denote the number of MIMO vector channel uses. Since each vector channel use carries one complex symbol per transmit antenna, the semantic encoder maps the source image $\bm{x}$ to a codeword $\bm{z} \in \mathbb{C}^{N_{\text{t}} T}$, which can be expressed as
\begin{equation}
  \bm{z} = \mathcal{E}_\theta(\bm{x}),
\end{equation}
where $\mathcal{E}_\theta(\cdot)$ denotes the semantic encoder parameterized by $\theta$. Accordingly, the bandwidth compression ratio (BCR), measured in transmitted complex channel symbols per source sample, is defined as $\mathrm{BCR} = N_{\text{t}}T/N$. The encoded signal is power-normalized to satisfy the per-stream transmit power constraint; for notational convenience, we still use $\bm{z}$ to denote the signal after normalization. The codeword $\bm{z}$ is reshaped into transmit vectors $\{\bm{s}_t\}_{t=1}^{T}$, where $\bm{s}_t \in \mathbb{C}^{N_{\text{t}}}$ collects the symbols sent from all transmit antennas at channel use $t$. The received signal at each channel use $t \in \{1, \ldots, T\}$ is
\begin{equation}
  \bm{y}_t = \bm{H} \bm{s}_t + \bm{n}_t,
\end{equation}
where $\bm{H} \in \mathbb{C}^{N_{\text{r}} \times N_{\text{t}}}$ is the channel matrix whose entries are independently and identically distributed as $h_{ij} \sim \mathcal{CN}(0,1)$, and $\bm{n}_t \sim \mathcal{CN}(\bm{0}, \sigma_{\text{ch}}^2 \bm{I}_{N_{\text{r}}})$ is the additive white Gaussian noise vector. We assume that the channel matrix $\bm{H}$ remains constant within each coherence block, i.e. slow fading. 

At the receiver, assuming perfect channel state information at the receiver (CSIR), $N_{\text{r}} \geq N_{\text{t}}$, and full column rank of $\bm{H}$, a zero-forcing (ZF) equalizer is applied to each received vector:
\begin{equation}
  \hat{\bm{s}}_t = (\bm{H}^{\mathsf{H}} \bm{H})^{-1} \bm{H}^{\mathsf{H}} \bm{y}_t,
\end{equation}
where $(\cdot)^{\mathsf{H}}$ denotes the conjugate transpose. The $T$ equalized vectors $\{\hat{\bm{s}}_t\}_{t=1}^{T}$ are concatenated into $\hat{\bm{z}} \in \mathbb{C}^{N_{\text{t}} T}$, which is then fed into the semantic decoder to reconstruct the source image:
\begin{equation}
  \hat{\bm{x}} = \mathcal{D}_{\phi}(\hat{\bm{z}}),
\end{equation}
where $\mathcal{D}_\phi(\cdot)$ denotes the semantic decoder parameterized by $\phi$ and $\hat{\bm{x}}$ is the reconstructed image.
\subsection{Eavesdropping Model}
\label{sec:eve-model}
We consider a passive eavesdropper, Eve, equipped with $N_{\text{e}}$ antennas, who attempts to recover the transmitted content through an independent MIMO fading channel. Specifically, at channel use $t$, Eve observes
\begin{equation}
  \bm{r}_t = \bm{G} \bm{s}_t + \bm{w}_t, \quad t \in \{1,\ldots,T\},
\end{equation}
where $\bm{G} \in \mathbb{C}^{N_{\text{e}} \times N_{\text{t}}}$ denotes the wiretap channel matrix and $\bm{w}_t \sim \mathcal{CN}(\bm{0}, \sigma_{\text{e}}^2 \bm{I}_{N_{\text{e}}})$ is the additive Gaussian noise at Eve. Following \cite{tang2025towards}, we consider a glass-box eavesdropper, which is a former legitimate user that can access the semantic encoder $\mathcal{E}_\theta$ but not the decoder $\mathcal{D}_\phi$. Moreover, to reflect a more practical setting \cite{Full-Duplex_Radio_for_Securing_Wireless_Network,Secure_MISO_Wiretap_Channels}, we further assume that Eve cannot obtain the instantaneous CSI of the Alice--Eve wiretap channel and only knows its statistical distribution. In general, Eve's inference can be expressed as
\begin{equation}
  \hat{\bm{x}}_{\mathrm e} = \mathcal{A}_{\psi}\bigl(\bm{r}, \mathcal{K}\bigr),
\end{equation}
where $\mathcal{A}_{\psi}$ denotes Eve's inference algorithm with parameters $\psi$, $\bm{r}=\{\bm{r}_t\}_{t=1}^{T}$ collects the received signals, and $\mathcal{K}$ is the auxiliary knowledge available to Eve, including the semantic encoder, semantic priors, and external generative models. Eve's goal is to produce an estimate $\hat{\bm{x}}_{\mathrm e}$ that is semantically close to the source image $\bm{x}$. Under these assumptions, Eve has to perform blind eavesdropping, and a typical approach is to jointly recover the source image and the wiretap channel through model inversion attacks, which can be expressed as
\begin{equation}
  (\hat{\bm{x}}_{\mathrm e},\hat{\bm{G}}_{\mathrm e})=\arg\min_{\tilde{\bm{x}},\tilde{\bm{G}}}\!\sum_{t=1}^{T}\bigl\|\tilde{\bm{G}}\,\bm{s}_t(\tilde{\bm{x}}) - \bm{r}_t\bigr\|_2^2 + \lambda_{\mathrm{tv}}\,\mathrm{TV}(\tilde{\bm{x}}),
  \label{eq:joint-inversion}
\end{equation}
where $\tilde{\bm{x}}\in\mathbb{R}^{3\times N_{\text{H}}\times N_{\text{W}}}$ and $\tilde{\bm{G}}\in\mathbb{C}^{N_{\text{e}}\times N_{\text{t}}}$ denote Eve's decision variables for the source image and the wiretap channel, respectively, $\bm{s}_t(\tilde{\bm{x}})$ denotes the $t$-th transmit vector obtained by passing $\tilde{\bm{x}}$ through the semantic encoder $\mathcal{E}_\theta$, $\mathrm{TV}(\cdot)$ denotes the total-variation regularizer, and $\lambda_{\mathrm{tv}}>0$ is its weight. However, solving \eqref{eq:joint-inversion} is challenging because the image and channel variables are strongly coupled in the bilinear forward model. Moreover, as the wiretap channel varies across realizations, a solver with a fixed configuration is too rigid to reliably recover coherent images. This limitation motivates the agentic design presented in Section~\ref{sec:agentic}.

\section{Agentic Eavesdropping}
\label{sec:agentic}
In this section, we present the proposed agentic eavesdropper, which reformulates the joint optimization in~\eqref{eq:joint-inversion} as a sequential decision process driven by an LLM agent. We first present the framework overview, and then describe the role of each agent and its tools in detail.

\subsection{Framework Overview}
\label{sec:agentic-overview}

As shown in Fig.~\ref{fig:agentic-eavesdropping}, the proposed agentic eavesdropper is organized around three agents: an optimization agent, a perception agent, and a refinement agent. These agents work together in a sequential decision-making framework to reconstruct the source image from the intercepted signal. Unlike conventional glass-box MIA using a single fixed-configuration solver, our framework decomposes the attack into a closed-loop process, where optimization produces candidates, perception evaluates them, and refinement improves promising results. The role of each agent is listed below.
\begin{itemize}
  \item \textbf{Optimization Agent:} Starting from an initial image and channel estimate, this agent performs short-step inversion to reduce the mismatch between the predicted and intercepted signals, while saving intermediate snapshots as candidates. If the result degrades or stalls, it can roll back to a previous checkpoint, update the channel estimate, or restart from a new initialization.

  \item \textbf{Perception Agent:} This agent evaluates the intermediate snapshots using image quality assessment tools and a multimodal LLM, judging whether each candidate is visually meaningful, e.g., whether it contains a coherent face, clear facial features, and obvious artifacts. This helps Eve select promising candidates rather than relying solely on the signal-level mismatch.

  \item \textbf{Refinement Agent:} This agent further improves promising but degraded candidates by applying a generative restoration tool under a VLM-grounded prompt. The refined candidate is then returned to the optimization agent as a warm start for codeword re-anchoring, ensuring consistency with the intercepted signal.
\end{itemize}
We then introduce the detailed workflow of each agent in the following.


\subsection{Optimization Agent}
\label{sec:optimization-agent}

The optimization agent solves the joint inverse problem in~\eqref{eq:joint-inversion}. To handle the strong coupling between the image estimate and the channel estimate, we provide the optimization agent with three update modes. Let $\mathcal{L}(\tilde{\bm{x}},\tilde{\bm{G}})$ denote the objective in~\eqref{eq:joint-inversion} at the current optimization state, where $\tilde{\bm{x}}$ is the image-side optimization variable. The agent can choose one of the following modes according to the current reconstruction status:
\begin{itemize}
  \item \textbf{Image-only:} The agent updates the image variable $\tilde{\bm{x}}$ while keeping the channel estimate $\tilde{\bm{G}}$ fixed:
  \begin{equation}
    \tilde{\bm{x}} \leftarrow \tilde{\bm{x}} - \eta_{\tilde{\bm{x}}}\,\nabla_{\tilde{\bm{x}}}\mathcal{L}(\tilde{\bm{x}},\tilde{\bm{G}}).
    \label{eq:upd-image}
  \end{equation}
  This mode is used when the current channel estimate is reliable and the image needs further refinement.

  \item \textbf{Channel-only:} The agent updates the channel estimate $\tilde{\bm{G}}$ while keeping the image variable $\tilde{\bm{x}}$ fixed:
  \begin{equation}
    \tilde{\bm{G}} \leftarrow \tilde{\bm{G}} - \eta_{\tilde{\bm{G}}}\,\nabla_{\tilde{\bm{G}}}\mathcal{L}(\tilde{\bm{x}},\tilde{\bm{G}}).
    \label{eq:upd-channel}
  \end{equation}
  This mode is useful as a short warm-up when the channel estimate is poorly initialized.

  \item \textbf{Joint:} The agent updates both $\tilde{\bm{x}}$ and $\tilde{\bm{G}}$:
  \begin{equation}
    \begin{aligned}
      \tilde{\bm{x}} &\leftarrow \tilde{\bm{x}} - \eta_{\tilde{\bm{x}}}\,\nabla_{\tilde{\bm{x}}}\mathcal{L}(\tilde{\bm{x}},\tilde{\bm{G}}), \\
      \tilde{\bm{G}} &\leftarrow \tilde{\bm{G}} - \eta_{\tilde{\bm{G}}}\,\nabla_{\tilde{\bm{G}}}\mathcal{L}(\tilde{\bm{x}},\tilde{\bm{G}}).
    \end{aligned}
    \label{eq:upd-joint}
  \end{equation}
  In practice, we use a smaller learning rate for $\tilde{\bm{G}}$ than for $\tilde{\bm{x}}$, since the channel estimate is more sensitive to unstable updates.
\end{itemize}

The agent organizes the inversion process into a set of independent and resumable trajectories, referred to as \emph{sessions}. Each session represents one attempt to jointly recover the private content and the wiretap channel, and maintains its own latent variables, channel estimate, optimizer states, checkpoints, and loss history. Instead of solving~\eqref{eq:joint-inversion} in a single long run, the agent advances each session through short updates, typically 20 to 80 steps per call, so that intermediate candidates can be observed and assessed before the next action is chosen. Based on the signal residual and the feedback from the perception agent, the optimization agent can choose among four actions for the active session. If the candidate remains plausible and the residual continues to decrease, the agent \emph{continues} the current session. If the residual stalls while the candidate is still plausible, it \emph{switches} to another update mode, e.g., updating the recovered content, the channel estimate, or both. If the candidate has recently degraded, it \emph{rolls back} to a previous checkpoint. If the session becomes stagnant and the candidate remains implausible, the agent \emph{terminates} the session and starts a new branch.

To escape poor basins caused by the bilinear coupling between the recovered content and the unknown wiretap channel, the agent further supports session-level branching inspired by the tree-of-thoughts paradigm~\cite{yao2023tree}. Specifically, a session is marked as stagnant when the relative loss variation within a sliding window falls below a threshold $\epsilon$ and no clear improvement is observed. When this condition is also supported by the perception-agent feedback, the current branch is stopped and a new session is initialized with a fresh channel estimate. In this way, the inversion is transformed from a single optimization trajectory into a tree-like search, where poor branches are pruned and promising ones are expanded. The saved checkpoints form a shared candidate pool, which serves as the hand-off interface for perception assessment and generative refinement.
\subsection{Perception Agent}
\label{sec:perception-agent}

A low loss in~\eqref{eq:joint-inversion} does not always indicate a good recovery. Since Eve has no access to the original source image, reference-based metrics such as PSNR, SSIM, or LPIPS cannot be used during the eavesdropping attack. Motivated by \cite{4kagent}, we introduce a perception mixture-of-experts (MoE) module that combines complementary experts to jointly evaluate the visual quality of the recovered image. Specifically, the perception MoE consists of two types of experts. The first type of experts are no-reference image quality assessment (NR-IQA) tools, which provide a fast and gradient-free quality estimate. However, NR-IQA scores alone cannot indicate whether the recovered image is semantically meaningful, especially when the MIA results are severely corrupted. To overcome this limitation, motivated by recent advances that leverage LLMs as judges~\cite{llm-as-a-judge} and by the strong visual understanding capabilities of multimodal LLMs~\cite{depictqa_v1}, the second type of expert is a multimodal LLM, which evaluates the semantic coherence of the recovered image. The outputs of the two types of experts are then fused into a unified quality signal that drives the perception agent's feedback.

Specifically, for every snapshot $\tilde{\bm{x}}$, the agent first computes the NR-IQA scores, given by
\begin{equation}
\bm{q}(\tilde{\bm{x}}) =
\big[
q_1(\tilde{\bm{x}}), q_2(\tilde{\bm{x}}), \ldots, q_M(\tilde{\bm{x}})
\big],
\label{eq:nr-iqa}
\end{equation}
where $M$ denotes the number of NR-IQA metrics, and each $q_m(\cdot)$ denotes one no-reference quality metric, such as NIQE for general naturalness, TOPIQ-NR for learned perceptual quality, and, when available, task-specific NR-IQA models, e.g., face-aware quality scores for facial images.

Subsequently, the agent feeds the snapshot $\tilde{\bm{x}}$ to the multimodal LLM to obtain structured visual evidence, given by
\begin{equation}
\mathcal{V}(\tilde{\bm{x}}) =
\mathrm{LLM}_{\mathrm{assess}}(\tilde{\bm{x}}, \mathcal{P}_{\mathrm{assess}}),
\label{eq:llm-assess}
\end{equation}
where $\mathcal{P}_{\mathrm{assess}}$ is a structured assessment prompt. Rather than asking the LLM to rank candidates or select the best reconstruction, this prompt asks it to inspect each noisy face reconstruction and return concrete visual evidence in a fixed JSON format. In our implementation, $\mathcal{V}(\tilde{\bm{x}})$ includes fields such as whether exactly one human face is visible, the estimated face pose, whether the facial components are complete, whether obvious artifacts are present, artifact descriptions, and a confidence value. It may also record auxiliary attributes, such as glasses, background type, hairstyle, hair color, and age range, which help describe the candidate but are not directly used in the scalar evidence score. The agent then converts the selected assessment fields into a visual feedback score, denoted by $e(\mathcal{V}(\tilde{\bm{x}}))$, which rewards reconstructions with recognizable content, complete key components, few artifacts, coherent structure, and high LLM confidence. The resulting score is fused with the no-reference IQA score and the observation-domain reconstruction error for candidate ranking. These two scores $\bm{q}(\tilde{\bm{x}})$ and $e(\mathcal{V}(\tilde{\bm{x}}))$ are then fed back to the optimization agent, which uses them together with the residual to decide the next session-level action, as described in Section~\ref{sec:optimization-agent}.
\begin{figure*}[!t]
  \centering
  \includegraphics[width=0.76\textwidth]{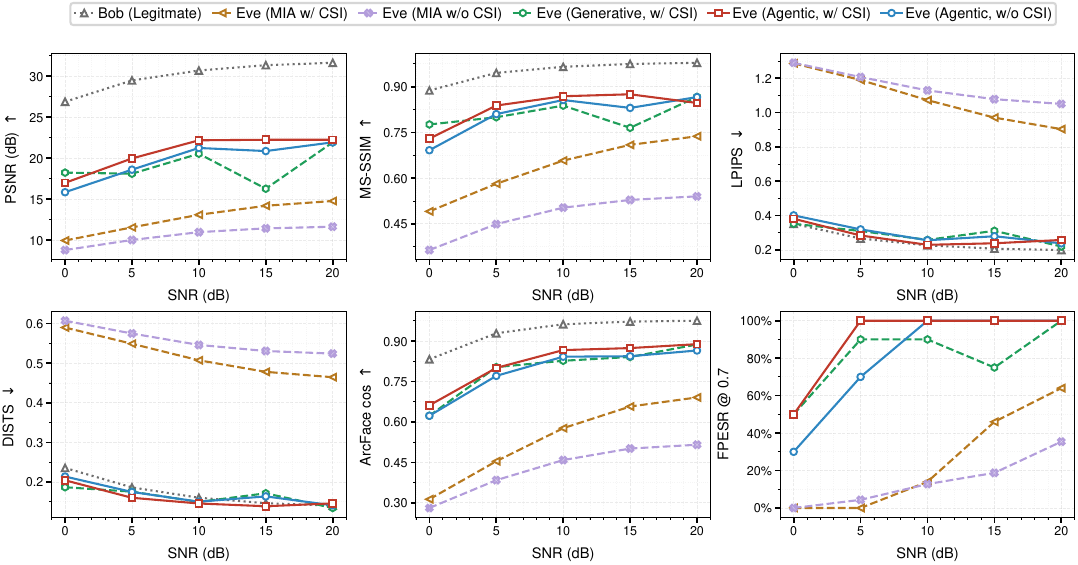}
  \caption{Performance comparison of the proposed agentic eavesdropper with baseline methods in terms of PSNR, MS-SSIM, LPIPS, DISTS, face cosine similarity, and FPESR, where SNR varies from $0$~dB to $20$~dB.}
  \label{fig:main_results}
\end{figure*}
\subsection{Refinement Agent}
\label{sec:refinement-agent}
The optimization and perception agents can guide the inversion toward snapshots that contain useful semantic cues. However, due to the ill-posed nature of MIA and the unknown wiretap channel, these snapshots may still exhibit blurred details, color distortion, or mild artifacts, and further gradient steps in the same session usually yield only marginal improvement. To overcome this limitation, we introduce the refinement agent, which uses an external generative prior to refine such a snapshot into a photorealistic candidate while keeping the result consistent with the intercepted signal.

Specifically, the refinement agent first queries the multimodal LLM to obtain a structured attribute description of the selected candidate snapshot $\tilde{\bm{x}}_{\mathrm{c}}$, which can be expressed as
\begin{equation}
\mathcal{C}(\tilde{\bm{x}}_{\mathrm{c}}) = \mathrm{LLM}_{\mathrm{describe}}(\tilde{\bm{x}}_{\mathrm{c}}, \mathcal{P}_{\mathrm{describe}}),
\label{eq:vlm-describe}
\end{equation}
where $\mathcal{P}_{\mathrm{describe}}$ denotes a captioning prompt that asks the LLM to record concrete visual attributes of the snapshot, such as object identity cues, appearance, pose, lighting, background, and remaining image-quality issues. We note that this description step is important because the refinement agent is not intended to perform blind restoration. Instead, it first extracts the visual evidence already present in the distorted snapshot and then uses this evidence to guide the subsequent restoration. Based on $\mathcal{C}(\tilde{\bm{x}}_{\mathrm{c}})$, the LLM then composes a restoration prompt $p_{\mathrm{rest}}$ that frames the task as restoration rather than imagination and instructs the model not to add new details that are not visible in the snapshot.

Conditioned on $\tilde{\bm{x}}_{\mathrm{c}}$ as the visual reference and on $p_{\mathrm{rest}}$ as the textual instruction, an external generative model produces a clean candidate, given by
\begin{equation}
\tilde{\bm{x}}_{\mathrm{g}} = \mathcal{R}_{\mathrm{gen}}(\tilde{\bm{x}}_{\mathrm{c}}, p_{\mathrm{rest}}),
\label{eq:gen}
\end{equation}
where $\mathcal{R}_{\mathrm{gen}}(\cdot)$ denotes the reference-conditioned image generator. We note that $\tilde{\bm{x}}_{\mathrm{g}}$ is expected to be more photorealistic than the distorted snapshot. However, $\tilde{\bm{x}}_{\mathrm{g}}$ cannot be directly accepted as the final recovery, since the generator does not observe the intercepted signal $\bm{r}$ during synthesis.

To address this, the refinement agent does not directly commit $\tilde{\bm{x}}_{\mathrm{g}}$ as the final answer. Instead, it hands $\tilde{\bm{x}}_{\mathrm{g}}$ back to the optimization agent, which warm-starts a fresh session from $\tilde{\bm{x}}_{\mathrm{g}}$ and runs a short joint-update burst against $\bm{r}$. Specifically, if the residual remains low and the perception agent confirms that the reconstruction is still visually meaningful, $\tilde{\bm{x}}_{\mathrm{g}}$ is admitted to the candidate pool as a refined candidate. Otherwise, $\tilde{\bm{x}}_{\mathrm{g}}$ is treated as a hallucination and discarded, and the corresponding session is rolled back to the pre-generation snapshot. In this way, the generative prior is used only as a guide that pulls promising snapshots toward natural images, while the consistency with $\bm{r}$ is enforced by the session-level mechanism described in Section~\ref{sec:optimization-agent}, which prevents the refinement agent from fabricating content that Alice did not transmit.


\section{Simulation Results}
\label{sec:simulation}
\subsection{Simulation Setup}
We evaluate the proposed agentic eavesdropper on the FFHQ dataset over a $2\times2$ MIMO Rayleigh fading channel with $\mathrm{SNR}\in[0,20]$~dB. The SemCom system is implemented by a DeepJSCC encoder--decoder pair with $\mathrm{BCR}=1/12$. In the proposed framework, Claude Sonnet 4.6 is used as the LLM agent, Gemini 2.5 Flash is used for visual assessment and captioning, and Google Nano Banana 2 is used for generative refinement. Each optimization session uses Adam with learning rates $5\times10^{-2}$ and $10^{-2}$ for the content latent and channel estimate, respectively, a TV weight of $5\times10^{-4}$, and $40$ steps per call. The NR-IQA score in~\eqref{eq:nr-iqa} is instantiated with TOPIQ-NR-Face and NIQE. We compare the proposed framework with three baselines: the legitimate receiver Bob, glass-box MIA with and without wiretap CSI, and an ablation variant that applies generative refinement with wiretap CSI but removes the agentic control loop. We report PSNR, MS-SSIM, LPIPS, DISTS, ArcFace cosine similarity~\cite{ArcFace}, and face privacy eavesdropping success rate (FPESR). A recovery is counted as successful privacy leakage when the ArcFace similarity exceeds $0.7$, and FPESR denotes the leakage rate over test images~\cite{chen2025privacy}.


\subsection{Effectiveness of the Proposed Eavesdropping}
Fig.~\ref{fig:main_results} compares the proposed agentic eavesdropper with baseline methods over $\mathrm{SNR}\in[0,20]$~dB. From this figure, we can observe that the proposed eavesdropper achieves a significantly higher FPESR than the glass-box MIA. Specifically, even at $\mathrm{SNR}=0$~dB where both MIA baselines (with and without wiretap CSI) stay close to $0\%$, the proposed eavesdropper still attains an FPESR of $30\%$--$50\%$, and its face cosine similarity, LPIPS, and DISTS curves track Bob closely and stay far above the MIA baselines. Moreover, the agentic eavesdropper without CSI nearly matches its CSI-aware version and clearly outperforms MIA with CSI, which shows that the agentic control loop can reduce the dependence on wiretap CSI. Furthermore, although the generative-refinement ablation tracks the proposed scheme on face cosine similarity at high SNR, it is less stable on FPESR even with wiretap CSI. Specifically, for $\mathrm{SNR}\geq 10$~dB, the proposed eavesdropper achieves up to a $10\%$ FPESR gain over the generative-refinement ablation. This result validates the effectiveness of the proposed agentic eavesdropping framework.

\begin{figure}[!t]
  \centering
  \includegraphics[width=0.85\linewidth]{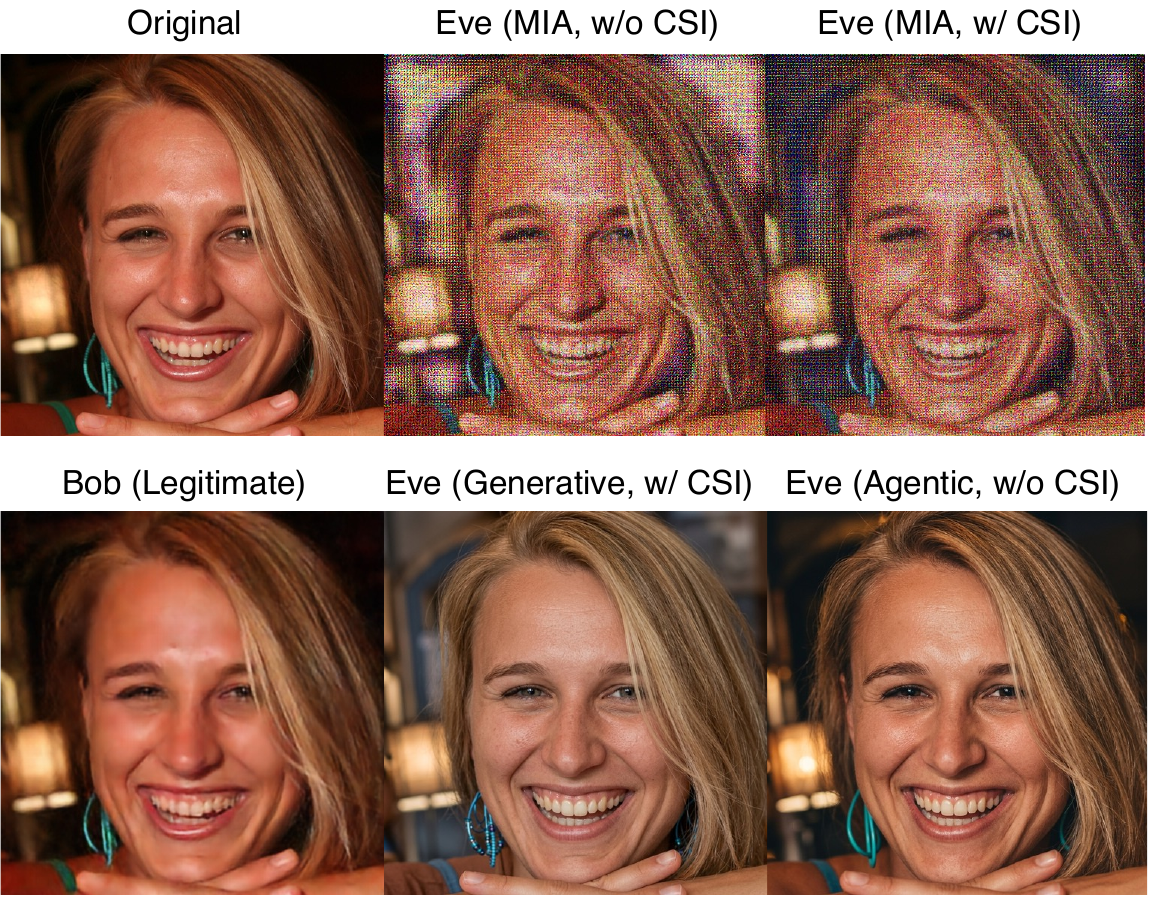}
  \caption{Visual comparison of reconstruction results obtained by Bob, conventional MIA, the generative-refinement ablation, and the proposed agentic eavesdropper at $\mathrm{SNR}=5$~dB.}
  \label{fig:vis}
\end{figure}

Fig.~\ref{fig:vis} visualizes the reconstruction results of different methods under $\mathrm{SNR}=5$~dB. From this figure, we can observe that the conventional MIA, with or without wiretap CSI, only recover noisy images in which facial components are severely distorted. In contrast, the generative-refinement ablation produces visually cleaner faces, but the recovered identity often deviates from the source image, which indicates that the generative prior introduces hallucinated content. Compared with these baselines, the proposed agentic eavesdropper consistently reconstructs faces that closely match the source image in identity, pose, and key attributes, even when the wiretap CSI is unavailable. These visual results further confirm the effectiveness of the proposed agentic eavesdropping framework.
\section{Conclusion}
\label{sec:conclusion}
In this paper, we have proposed an agentic eavesdropper to reveal the privacy risk of SemCom under a practical glass-box setting where the eavesdropper only knows the statistical distribution of the wiretap channel. Specifically, the proposed framework decomposes the attack into three cooperating agents, namely an optimization agent, a perception agent, and a refinement agent, which jointly enforce signal-level fidelity and semantic plausibility through session-level branching, rollback, and codeword re-anchoring. Simulation results on the FFHQ dataset demonstrated that the proposed eavesdropper achieves more than $75\%$ FPESR at $\mathrm{SNR}\geq 5$~dB even without wiretap CSI, exposing a non-negligible privacy threat that future secure SemCom designs must address.

\bibliographystyle{IEEEtran}
\bibliography{IEEEabrv,references}

\begin{thebibliography}{10}
\providecommand{\url}[1]{#1}
\csname url@samestyle\endcsname
\providecommand{\newblock}{\relax}
\providecommand{\bibinfo}[2]{#2}
\providecommand{\BIBentrySTDinterwordspacing}{\spaceskip=0pt\relax}
\providecommand{\BIBentryALTinterwordstretchfactor}{4}
\providecommand{\BIBentryALTinterwordspacing}{\spaceskip=\fontdimen2\font plus
\BIBentryALTinterwordstretchfactor\fontdimen3\font minus \fontdimen4\font\relax}
\providecommand{\BIBforeignlanguage}[2]{{%
\expandafter\ifx\csname l@#1\endcsname\relax
\typeout{** WARNING: IEEEtran.bst: No hyphenation pattern has been}%
\typeout{** loaded for the language `#1'. Using the pattern for}%
\typeout{** the default language instead.}%
\else
\language=\csname l@#1\endcsname
\fi
#2}}
\providecommand{\BIBdecl}{\relax}
\BIBdecl

\bibitem{Semantic1}
D.~G{\"{u}}nd{\"{u}}z, Z.~Qin, I.~E. Aguerri, H.~S. Dhillon, Z.~Yang, A.~Yener, K.~Wong, and C.~Chae, ``Beyond transmitting bits: Context, semantics, and task-oriented communications,'' \emph{{IEEE} J. Sel. Areas Commun.}, vol.~41, no.~1, pp. 5--41, 2023.

\bibitem{Semantic_security_zhaohui}
Z.~Yang, M.~Chen, G.~Li, Y.~Yang, and Z.~Zhang, ``Secure semantic communications: Fundamentals and challenges,'' \emph{{IEEE} Netw.}, vol.~38, no.~6, pp. 513--520, 2024.

\bibitem{Semantic_security_maojun}
M.~Zhang, Y.~Li, Z.~Zhang, G.~Zhu, and C.~Zhong, ``Wireless image transmission with semantic and security awareness,'' \emph{{IEEE} Wirel. Commun. Lett.}, vol.~12, no.~8, pp. 1389--1393, 2023.

\bibitem{Semantic_security_yuhao}
Y.~Chen, Q.~Yang, Z.~Shi, and J.~Chen, ``The model inversion eavesdropping attack in semantic communication systems,'' in \emph{IEEE Glob. Commun. Conf. (GLOBECOM)}, 2023, pp. 1--6.

\bibitem{tang2025towards}
S.~Tang, Y.~Chen, Q.~Yang, R.~Zhang, D.~Niyato, and Z.~Shi, ``Towards secure semantic communications in the presence of intelligent eavesdroppers,'' \emph{arXiv:2503.23103}, 2025.

\bibitem{Full-Duplex_Radio_for_Securing_Wireless_Network}
Y.~Hua, ``Advanced properties of full-duplex radio for securing wireless network,'' \emph{{IEEE} Trans. Signal Processing}, vol.~67, no.~1, pp. 120--135, 2019.

\bibitem{Secure_MISO_Wiretap_Channels}
H.-M. Wang, T.~Zheng, and P.~Mu, ``Secure miso wiretap channels with multi-antenna passive eavesdropper via artificial fast fading,'' in \emph{IEEE Commun. Conf. (ICC)}, 2014, pp. 5396--5401.

\bibitem{yao2023react}
S.~Yao, J.~Zhao, D.~Yu, N.~Du, I.~Shafran, K.~R. Narasimhan, and Y.~Cao, ``React: Synergizing reasoning and acting in language models,'' in \emph{Proc. Int. Conf. Learn. Repr. (ICLR)}, 2023.

\bibitem{yao2023tree}
S.~Yao, D.~Yu, J.~Zhao, I.~Shafran, T.~Griffiths, Y.~Cao, and K.~Narasimhan, ``Tree of thoughts: Deliberate problem solving with large language models,'' \emph{Proc. Adv. Neural Inf. Process. Syst. (NeurIPS)}, vol.~36, pp. 11\,809--11\,822, 2023.

\bibitem{llm-as-a-judge}
J.~Gu, X.~Jiang, Z.~Shi, H.~Tan, X.~Zhai, C.~Xu, W.~Li, Y.~Shen, S.~Ma, H.~Liu \emph{et~al.}, ``A survey on {LLM}-as-a-judge,'' \emph{The Innov.}, 2024.

\bibitem{zhang2025toward}
R.~Zhang, S.~Tang, Y.~Liu, D.~Niyato, Z.~Xiong, S.~Sun, S.~Mao, and Z.~Han, ``Toward agentic ai: Generative information retrieval inspired intelligent communications and networking,'' \emph{IEEE Commun. Mag.}, 2025.

\bibitem{4kagent}
Y.~Zuo, Q.~Zheng, M.~Wu, X.~Jiang, R.~Li, J.~Wang, Y.~Zhang, G.~Mai, L.~V. Wang, J.~Zou, X.~Wang, M.-H. Yang, and Z.~Tu, ``{4KAgent}: Agentic any image to {4K} super-resolution,'' \emph{arXiv preprint arXiv:2507.07105}, 2025.

\bibitem{depictqa_v1}
Z.~You, Z.~Li, J.~Gu, Z.~Yin, T.~Xue, and C.~Dong, ``Depicting beyond scores: Advancing image quality assessment through multi-modal language models,'' in \emph{Proc. Eur. Conf. Comput. Vis. (ECCV)}, 2024, pp. 259--276.

\bibitem{ArcFace}
J.~Deng, J.~Guo, N.~Xue, and S.~Zafeiriou, ``Arcface: Additive angular margin loss for deep face recognition,'' in \emph{Proc. IEEE/CVF Conf. Comput. Vis. Pattern Recog. (CVPR)}, 2019, pp. 4685--4694.

\bibitem{chen2025privacy}
W.~Chen, Q.~Yang, S.~Shao, S.~Tang, Z.~Shi, and S.~Yu, ``Privacy-preserving semantic communication over wiretap channels with learnable differential privacy,'' \emph{arXiv:2510.23274}, 2025.

\end{thebibliography}

\end{document}